\documentclass[letterpaper,12pt]{article}   

\usepackage{amsfonts}
\usepackage{graphicx}
\usepackage{epsfig}
\usepackage{graphics}
\usepackage{mathtools}
\usepackage{amsmath}
\usepackage{cite}

\makeatletter
\newcommand{\row}[1]%
{\mathord{\buildrel{\lower3pt%
\hbox{$\scriptscriptstyle\rightarrow$}}\over #1}}

\newcommand{\dyadic}[1]{\mathord{\dyadic@rrow{#1}}}
\newcommand{\dyadic@rrow}[1]{
\begin{picture}(12,12)(-1,0)
\put(-3,12){\makebox(0,0)[t]{$\scriptscriptstyle\downarrow$}}
\put(-3,13){\makebox(0,0)[l]{$\scriptscriptstyle\longrightarrow$}}
\put(5,0){\makebox(0,0)[b]{$#1$}}
\end{picture}
}

\newcommand{\bra}[1]{\bigl\langle #1 \bigr|}
\newcommand{\ket}[1]{\bigl| #1 \bigr\rangle}

\begin{document}

\begin{center}
{\large Teleportation with Multiple Accelerated Partners}

\vspace{0.5cm}
\renewcommand{\thefootnote}{\fnsymbol{footnote}}
 Alaa Sagheer $^{\dagger \ddagger} $ and Hala Hamdoun$^\ddagger$\\
$^\dagger$Department  of Mathematics\\
$^\ddagger$Center for Artificial Intelligence and RObotics (CAIRO)\\
Faculty of Science, Aswan University, Aswan,
Egypt\\
Email: asagheer@aswu.edu.eg\\
           hala@cairo.aswu.edu.eg\\

\end{center}
{\bf Abstract}:  As the current revolution  in communication is underway, quantum teleportation can increase the level of security in quantum communication applications. In this paper, we present a quantum teleportation procedure that capable to teleport either accelerated or non-accelerated information through different quantum channels. These quantum channels are based on accelerated multi-qubit states, where each qubit of each of these channels represent a partner. Namely, these states are the the W state, Greenberger-Horne-Zeilinger (GHZ) state, and the GHZ-like state.
Here, we show that the fidelity of teleporting accelerated information
 is higher than the fidelity of teleporting non-accelerated information, both through a quantum channel that is based on accelerated state. Also, the comparison among the performance of these three channels shows that the degree of  fidelity depends on  type of the used channel, type of the measurement, and value of the acceleration.  The result of comparison concludes that teleporting information through channel that is based on the GHZ state is more robust than teleporting information through channels that are based on the other two states.
 For future work, the proposed procedure can be generalized later to achieve communication through a wider quantum network.

\textbf{keywords:} quantum information, quantum communication, teleportation, multi-qubit channels, accelerated information, fidelity

\section{Introduction}
 It is known that, entanglement is a fundamental resource for many of quantum information processing (QIP) themes
 \cite{Horodecki}, such as quantum cryptography \cite{Bennett1}, quantum computation \cite{RIEFFEL},
 and quantum communication\cite{Metwally2004,Metwally2009}. One of the exciting applications of quantum communication, which are based on entanglement, is quantum teleportation. Recently, quantum teleportation has been paid much attention both theoretically and experimentally since it could make quantum communication essentially instant \cite{Ursin}. Most of the current quantum teleportation procedures achieve teleportation of non-accelerated information through non-accelerated states \cite{Gorbachev,Joo,Kan}. With the rapid development in communication domain, there is an urgent need to develop new procedures achieve teleportation of either accelerated or non-accelerated information through channel based on accelerated multi qubit entangled states with a high level of security and efficiency.

Since 20 years ago, the first teleportation protocol which uses two qubit channel is presented theoretically
  by Bennett et al. \cite{Bennett2}. Next, several protocols of quantum teleportation based on Bennet protocol
   have been developed, some of them are realized in experiments \cite{Zeilinger,Zeilinger2,Ivan}.
   All of these protocols teleported unknown information from the sender to a remote receiver,
    where both of them are spatially separated via a classical channel.
    According to Bell basis measurements performed by the sender, the receiver applies the corresponding
    unitary operations on his single qubit and obtains the original information with certainty.

Next, the quantum teleportation with multi-qubit systems have been
attracted much attention due to its generalization
  of the previous teleportation procedures with two qubit systems.
  The main difference between quantum teleportation of
  two qubit stystems and that of three qubit systems
  is the existence of another receiver, who contribute to teleport the state from the sender to the reserver
  \cite{Jeongwan}. In most of these protocols, quantum teleportation with multi-qubit
   is achieved through non-accelerated states. Karlsson et al. \cite{Karlsson},
   for example, demonstrated a teleportation with three-qubit channel capable to teleport
    unknown information from the sender to any of the two receivers. In Karlsson protocol,
     only one of the two receivers can fully reconstruct the teleported information
     conditioned on the measurement result of the other reciever.
     Later, Alsing et al. introduced the first  protocol achieved teleportation through
 uniformaly accelerated state \cite{Alsing2,Alsing1}. They described the process of teleportation between the sender, who is not accelerated, and the reciver who is in a uniform acceleration with respect to the sender.
Recently, Metwally discussed the possibility of using maximum and
partial entangled qubits to perform accelerated quantum
teleportation with accelerated or non-accelerated information
\cite{Metwally1,Metwally2}.

 In this paper, we propose a quantum  protocol achieves teleportation of either accelerated or non-accelerated information through channel based on accelerated multi-qubit state. The proposed protocol differs from other protocols in that all qubits of  the used channels, here, are accelerated. In addition, we investigate here the behaviour of the teleported information, either accelerated or non-accelerated, through three kinds of channels and conduct a comparison among them. Theses channels are based on the W state, the Greenberger-Horne-Zeilinger (GHZ) state,  and the GHZ-like state.

  The proposed protocol teleports the information from  the sender (Alice) to the receiver (Bob) with the help of the third qubit (Charlie) according to the following scenario. First, the sender Alice performs Bell basis measurements on her two qubits, one is the information qubit and the other is the qubit entangled to other qubits. Second, she sends the measurement result to both the receiver Bob and the receiver Charlie. Third, the receiver Charlie performs a single qubit measurement according to  Alice measurement result and, then, sends the measurement result to  Bob. Finally, the receiver Bob can retrieve the teleported information.

 The paper is organized as follows: \textbf{Section 2} describes the accelerated states use as a quantum channels.   The  teleportation procedure through the three channels are provided and discussed in \textbf{Section 3}, where the fidelity of each channel is investigated. Finally, \textbf{Section 4} concludes the paper and shows our future work.

\section{Quantum Channels of Teleportation}
In this paper, we adopt three different communication channels that are based on multi qubit accelerated state, where each qubit of each of these channels represents a partner who moves in uniform acceleration. These accelerated states can be described as an entangled qubit of two modes monochromatic in with frequency $\omega_k$:
\begin{equation}\label{rindler}
\ket{0_k}_M=\cos r\ket{0_k^{+}}_{I}\ket{0_k^{-}}_{II}+ \sin r \ket{1_k^{+}}_{I}\ket{1_k^{-}}_{II}, \quad
\ket{1_k}_M= \ket{1_k^{+}}_{I}\ket{0_k^{-}}_{II},
\end{equation}
where $k=0,1,2$ and $3$ for teleported information qubit, first, second and third qubit of channel, respectively.
The definition of the acceleration for one qubit with respect to the observer, $r$, is given by $\tan r_k= exp[ -\pi \omega_k
c/a_k]$ where $0 \leq  r_k \leq \pi/4 $. "$a_k$" is the acceleration of the accelerated qubits with respect to the speed of light, where $0 \leq a_k \leq \infty $. "$\omega_k$" is
the frequency of the traveling qubit and "$c$" is the speed of
light. $\ket{n}_I$ and $\ket{n}_{II}$, (n = 0, 1), indicate two causally disconnected regions in the Rindler space \cite{Martın}.

According to Eq.(\ref{rindler}), the density operator of the accelerated states is described by the following 8X8 matrix:

\begin{equation}
\rho_{state}^I=\left(
\begin{matrix}
\varepsilon_{11} & \varepsilon_{12} & \varepsilon_{13} & \varepsilon_{14} & \varepsilon_{15} & \varepsilon_{16} & \varepsilon_{17} & \varepsilon_{18} \\
\varepsilon_{21} & \varepsilon_{22} & \varepsilon_{23} & \varepsilon_{24}& \varepsilon_{25} & \varepsilon_{26} & \varepsilon_{27} & \varepsilon_{28} \\
\varepsilon_{31} & \varepsilon_{32} & \varepsilon_{33} & \varepsilon_{34}&\varepsilon_{35} & \varepsilon_{36} & \varepsilon_{37} & \varepsilon_{38} \\
\varepsilon_{41} & \varepsilon_{42} & \varepsilon_{43} & \varepsilon_{44} & \varepsilon_{45} & \varepsilon_{46} & \varepsilon_{47} & \varepsilon_{48}\\
\varepsilon_{51} & \varepsilon_{52} & \varepsilon_{53} & \varepsilon_{54} & \varepsilon_{55} & \varepsilon_{56} & \varepsilon_{57} & \varepsilon_{58}\\
\varepsilon_{61} & \varepsilon_{62} & \varepsilon_{63} & \varepsilon_{64} & \varepsilon_{65} & \varepsilon_{66} & \varepsilon_{67} & \varepsilon_{68}\\
\varepsilon_{71} & \varepsilon_{72} & \varepsilon_{73} & \varepsilon_{74} & \varepsilon_{75} & \varepsilon_{76} & \varepsilon_{77} & \varepsilon_{78}\\
\varepsilon_{81} & \varepsilon_{82} & \varepsilon_{83} & \varepsilon_{84} & \varepsilon_{85} & \varepsilon_{86} & \varepsilon_{87} & \varepsilon_{88}\\
\end{matrix}
\right)
\end{equation}

The elements of this matrix, in the region $I$, depend on the current state by the same way as given in \cite{Alaa}.

\begin{itemize}
  \item In case of the W-state with a probability equals $\frac{1}{3}$, we have the following elements:
  \begin{eqnarray}\label{W}
\varepsilon_{22}&=&C_2^2C_3^2, \quad  \varepsilon_{23}=
\varepsilon_{32}= C_1C_2C_3^2, \quad\varepsilon_{25} =
\varepsilon_{52} =C_1C_3C_2^2
\nonumber\\
\varepsilon_{33}&=& C_1^2C_3^2 \quad \varepsilon_{35}=\varepsilon_{53}= C_2C_3C_1^2
\nonumber\\
\varepsilon_{44}&=& S_2^2C_3^2+S_1^2C_3^2 \quad
\varepsilon_{46}=\varepsilon_{64}=C_2C_3S_1^2\quad
\varepsilon_{47}=\varepsilon_{74}=C_2C_3S_1^2
\nonumber\\
\varepsilon_{55} &=& C_1^2C_2^2, \quad
\varepsilon_{66}= S_1^2C_2^2+S_3^2C_2^2, \quad \varepsilon_{67} = \varepsilon_{76}= C_1C_2S_3^2
\nonumber\\
\varepsilon_{77}&=& S_1^2C_1^2+S_3^2C_1^2, \quad  \varepsilon_{88}= S_1^2S_2^2+S_1^2S_3^2+S_2^2S_3^2
\end{eqnarray}

  \item In case of the GHZ-state with a probability equals $\frac{1}{2}$, we have the following elements:
  \begin{eqnarray}\label{G}
\varepsilon_{11}&=&C_1^2C_2^2C_3^2, \quad  \varepsilon_{22}= C_2^2C_3^2S_1^2, \quad\varepsilon_{33} =C_1^2C_3^2S_2^2
\nonumber\\
\varepsilon_{44}&=& C_3^2S_1^2S_2^2, \quad \varepsilon_{55}=C_1^2C_2^2S_3^2, \quad \varepsilon_{66}=C_2^2S_1^2S_3^2
\nonumber\\
\varepsilon_{77}&=& C_1^2S_2^2S_3^2 \quad
\varepsilon_{88}=S_1^2S_2^2S_3^2+1 \quad
\varepsilon_{81}=\varepsilon_{18}=C_1C_2C_3
\end{eqnarray}

  \item In case of the GHZ-like state with a probability equals $\frac{1}{4}$, we have the following elements:

   \begin{eqnarray}\label{GL}
\varepsilon_{22}&=&C_2^2C_3^2, \quad  \varepsilon_{25}= \varepsilon_{52}= C_1C_3C_2^2, \quad \varepsilon_{23}= \varepsilon_{32}= C_1C_2C_3^2
\nonumber\\
\varepsilon_{28}&=&\varepsilon_{82}= C_2C_3, \quad \varepsilon_{33}=C_1^2C_3^2, \quad \varepsilon_{35}=\varepsilon_{53}=C_2C_3C_1^2
\nonumber\\
\varepsilon_{38}&=&\varepsilon_{38}= C_1C_3 \quad
\varepsilon_{44}=S_2^2C_3^2+S_1^2C_3^2 \quad
\varepsilon_{47}=\varepsilon_{74}=C_1C_3S_2^2
\nonumber\\
\varepsilon_{46}&=&\varepsilon_{64}= C_2C_3S_1^2 \quad
\varepsilon_{55}=C_1^2C_2^2 \quad
\varepsilon_{58}=\varepsilon_{85}=C_2C_2C_3 \quad
\nonumber\\
\varepsilon_{66}&=&S_1^2C_2^2+S_3^2C_2^2 \quad
\varepsilon_{67}=\varepsilon_{76}=C_1C_2S_3^2 \quad
\varepsilon_{77}=S_2^2C_1^2+S_3^2C_1^2
\nonumber\\
\varepsilon_{88}&=&S_1^2S_2^2+S_1^2S_3^2+S_2^2S_3^2+1
\end{eqnarray}
\end{itemize}

where $ C_k=\cos r_k, S_k=\sin r_k$ and $k=1,2,3$ are  first,  second and  third qubit (or partner) of the corresponding channel, respectively.
In the following section, we use these three accelerated states to teleport either accelerated or non-accelerated information.

\section{The proposed procedure of teleportation }
The procedure starts with teleporting the information from a sender (Alice) to a receiver (Bob) with assistance of the third quibt, or a second receiver (Charlie).

Suppose that the information which Alice wishes to teleport to Bob is coded in the state $\ket{\psi}_0$, where
\begin{eqnarray}\label{target}
\ket{\psi}_0&=&\alpha\ket{0}_0+\beta\ket{1}_0
\end{eqnarray}
where $\alpha$ and $\beta$ are two complex numbers satisfying
$|\alpha|^2$ + $|\beta|^2$ = 1.

In order to teleport the $\ket{\psi}_0$, the partners follow the following steps:

{\bf Step 1}: Alice combines the teleported information state with her  qubit of the accelerated entangled state.

{\bf Step 2}: Alice performs Bell Measurements (BM) on her two
qubits. These measurements are described by:
\begin{eqnarray}\label{bell}
\rho_{{\psi}}^{\pm}&=&\frac{1}{2}
\Bigl(\ket{00}\bra{00}\pm\ket{00}\bra{11}\pm\ket{11}\bra{00}+
\ket{11}\bra{11}\Bigr)
\nonumber\\
\rho_{{\phi}}^{\pm}&=&\frac{1}{2}
\Bigl(\ket{01}\bra{01}\pm\ket{01}\bra{10}\pm\ket{10}\bra{01}+
\ket{10}\bra{10}\Bigr)
\end{eqnarray}

{\bf Step 3}: Charlie makes the Von Neuman measurement on his qubit . Then, he and Alice send their measurements to Bob.

{\bf Step 4} Based on Alice and Charlie measurements, Bob does one of the appropriate unitary operation, bit-flip (X), phase flip (Z) or bit-phase flip (Y) qubit operation, to get the initial teleported information.

\begin{figure}[h]
    \centering
    \includegraphics[width=25pc,height=15pc]{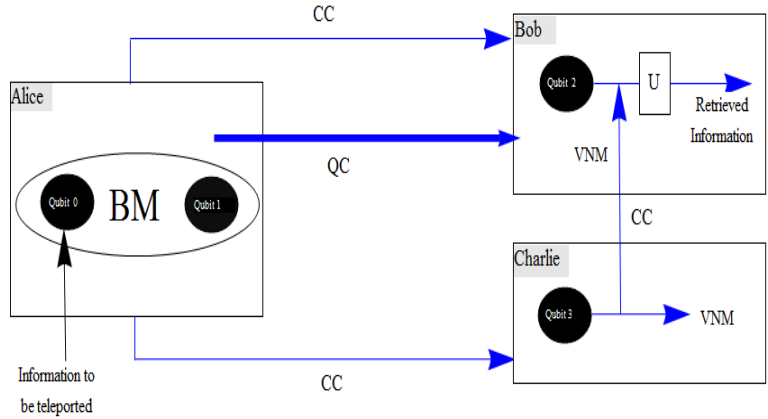}
    \caption{A Schematic diagram of the teleportation process.}
    \label{telep}
\end{figure}

A schematic diagram of the proposed teleportation procedure is depicted in Figure \ref{telep}. Qubit 0 denotes the qubit which contains the coded information that will be teleported and qubit 1,2 and 3 denote the three qubits of quantum channel (QC) that belong to Alice, Bob, and Charlie, respectively. Alice performs the appropriate Bell measurement on qubit 0 and qubit 1, then she informs both Bob and Charlie about her measurement through a classical channel (CC). For the sake of assisting Alice and Bob, Charlie  makes a single qubit measurement (Von Neuman measurement, VNM) on his qubit 3 and, then, transmits his result to Bob across a classical communication channel. Finally, Bob performs an appropriate unitary operation (U) on qubit 2 in order to retrieve the teleported information.

In the paper, it is assumed that the teleported information may be
accelerated or non-accelerated. For the
non-accelerated  case, the information is coded in the following single qubit:
\begin{eqnarray}\label{target1}
\rho_{{\psi}_0}&=&\alpha^2 \ket{0}\bra{0}+\alpha \beta
\ket{0}\bra{1}+\alpha \beta \ket{1}\bra{0}+\beta^2\ket{1}\bra{1},
\end{eqnarray}
 On the other hand, the accelerated information is coded in the following single qubit:
\begin{eqnarray}\label{target2}
\rho_{{\psi}_0}^{(I)}&=&\alpha^2 C_0^2\ket{0}\bra{0}+\alpha \beta
C_0\ket{0}\bra{1}+\alpha \beta
C_0\ket{1}\bra{0}+(\alpha^2 S_0^2 +\beta^2)\ket{1}\bra{1}
\end{eqnarray}

Now, the channels are capable to send information from Alice to Bob according to the teleportation procedure which is described above. In the following, we  evaluate the fidelity of the teleported information via the three predefined channels.

\subsection{ The W-state as a quantum channel}
In this section, we will use the description of the accelerated W- state, given in Eq.(\ref{W}), in order to teleport the non-accelerated information which is given in Eq.(\ref{target1}). For example, if Alice measures $\rho_{{\psi}}^{+}$,
as on of Bell state measurements, the other two qubits are projected into  the
following density operator:
\begin{equation}
\rho^{tot}=\left(
\begin{matrix}
\varrho_{11} & \varrho_{12} & \varrho_{13} & \varrho_{14} \\
\varrho_{21} & \varrho_{22} & \varrho_{23} & \varrho_{24} \\
\varrho_{31} & \varrho_{32} & \varrho_{33} & \varrho_{34} \\
\varrho_{41} & \varrho_{42} & \varrho_{43} & \varrho_{44} \\
\end{matrix}
\right),
\end{equation}
where,

\begin{eqnarray}
\varrho_{11}&=&\beta^2 C_2^2C_3^2, \quad  \varrho_{12}=
\varrho_{21}= \alpha \beta C_1C_3C_2^2, \quad\varrho_{13} =
\varrho_{31} =\alpha \beta C_1C_2C_3^2
\nonumber\\
\varrho_{22}&=& \alpha^2 C_1^2C_2^2+\beta^2 C_2^2S_3^2+\beta^2
S_1^2C_2^2, \quad \varrho_{23}= \alpha^2 C_2C_3C_1^2+ \beta^2
C_2C_3S_1^2,
\nonumber\\
\varrho_{24}&=&\varrho_{42}=\alpha \beta C_1C_2^2S_3, \quad
\varrho_{32}=\alpha^2C_2C_3C_3C_1^2+\beta^2C_2C_3S_1^2,
\nonumber\\
\varrho_{33}&=&\alpha^2C_1^2C_3^2+
\beta^2C_3^2S_2^2+\beta^2C_3^2S_1^2, \quad
\varrho_{34} = \varrho_{43} =\alpha \beta C_1C_3S_2^2,
\nonumber\\
\varrho_{44}&=&
\alpha^2C_1^2(S_2^2+S_3^2)+\beta^2S_3^2(S_1^2+S_2^2)+\beta^2S_1^2S_2^2)
\end{eqnarray}
Now, Charlie's measurement deicides the success or the failure of
the teleportation process itself. If Charlie measures 1,
 the teleportation fails. If Charlie measures  0, Bob can do an appropriate operation, from those given in Table 1, to retrieve the teleported information according to Alice measurement information.

\begin{table}[h]
\caption {Bob Unitary Operation when Charlie measures $''0''$} \label{tab1}
\centering
\begin{tabular}{|c|c|c|}
\hline
  Alice measurement & Charlie measurement & Bob unitary operation \\
  \hline
  $\rho_{{\psi}}^{+}$ & {0}& X \\
  $\rho_{{\psi}}^{-}$ & {0} & Y \\
  $\rho_{{\phi}}^{+}$ & {0} & I \\
  $\rho_{{\phi}}^{-}$ & {0} & Z \\\hline
\end{tabular}
\end{table}

The fidelity of the retrieved information at Bob side is given
as:
\begin{equation}\label{Wnon}
\mathcal{F}_w^{na}=\alpha^4C_1^2C_3^2 + \alpha^2\beta^2C3^2(S2^2 + S1^2)
+ 2\alpha^2\beta^2C_1C_2C_3^2 + \beta^4C_2^2C_3^2
\end{equation}

On the other hand,  the accelerated W- state, given in Eq.(\ref{W}), is used to teleport the non-accelerated information which is given in Eq.(\ref{target2}) using the same proposed teleportation procedure..
In this case, Alice qubits are projected into one of the Bell state measurements, for example
$\rho_{{\phi}}^{+}$,  and the two other qubits are projected into
$4\times 4$ matrix, its elements are given as:

\begin{eqnarray}
{\varrho_{\bf11}}&=&\alpha^2 C_0^2C_2^2C_3^2, \quad \varrho_{12}=
\varrho_{21}=\alpha \beta C_0C_1C_3C_2^2, \quad\varrho_{13}=
\varrho_{31} =\alpha \beta C_0C_1C_2C_3^2
\nonumber\\
 \varrho_{22}&=& \alpha^2 C_0^2C_2^2S_3^2+\alpha^2
C_0^2C_2^2S_1^2+\alpha^2 C_1^2C_2^2S_0^2+ \beta^2 C_1^2C_2^2
\nonumber\\
  \varrho_{23}&=& \alpha^2 C_0^2C_2C_3S_1^2+\alpha^2 C_1^2C_2C_3S_0^2+ \beta^2 C_1^2C_2C_3
\nonumber\\
\varrho_{24}&=&\varrho_{42}=\alpha \beta C_0C_1C_2^2C_3,\quad
\varrho_{32}=\alpha^2C_1^2C_2C_3S_0^2+ \alpha^2
C_0^2C_2C_3S_1^2+\beta^2C_1^2C_2C_3
\nonumber\\
\varrho_{33}&=&\alpha^2C_0^2C_3^2S_2^2+
\alpha^2C_0^2C_3^2S_1^2+\alpha^2C_1^2C_3^2S_0^2+\beta^2C_1^2C_3^2,
\nonumber\\
\varrho_{34}&=& \varrho_{43} =\alpha \beta C_0C_1C_3S_2^2
\nonumber\\
\varrho_{44}&=&
\alpha^2C_0^2S_3^2(S_1^2+S_2^2)+\alpha^2C1_2S_0^2(S_2^2+S_3^2)
\nonumber\\
&&+\beta^2C_1^2(S_2^2+S_3^2)+\alpha^2C_0^2S_1^2S_2^2
\end{eqnarray}
Now , Bob can retrieve the teleportaed information using an appropriate operation, from those given in Table \ref{tab1}, with a fidelity given as:

\begin{eqnarray}\label{Wacc}
\mathcal{F}_w^{ac}&=&
\alpha^4C_3^2(C_0^4C_2^2+S_0^4C_1^2)+\beta^4C_1^2C_3^2+
\alpha^2\beta^2C_0^2C_3^2(S_1^2+S_2^2)
\nonumber\\
&&+\alpha^4S_0^2C_0^2C_3^2(S_1^2+S_2^2)+ 2 \alpha^2\beta^2C_0^2C_1C_2C_3^2+2
\alpha^2\beta^2S_0^2C_1^2C_3^2
\end{eqnarray}

Figure 2, describes the fidelities of the teleported  information
which is coded in the qubit forms stated in Eq.(\ref{target1}) and Eq.(\ref{target2}), for the
non-accelerated and accelerated information, respectively. The panel (a) in Figure 2 shows the behavior of the fidelity of both the accelerated information
$\mathcal{F}w^{ac}$ and the non-accelerated information
$\mathcal{F}^{na}_w$. It is clear that  at zero acceleration of
all  qubits, the fidelities are maximum. However,  the fidelities decrease to reach its
minimum bounds when the acceleration $r_i$ of channel qubits reaches 0.8. Also, we notice that the degradation rate
of $\mathcal{F}^{ac}_w$ is faster than that depicted for
$\mathcal{F}^{nac}_w$. This shows that teleporting non-accelerated information is  better
than teleporting accelerated information through the accelerated W-state.

The panel (b) in Figure 2 displays the behavior of $\mathcal{F}^{ac}_w$, where it is
assumed that the teleported information is accelerated with different values of
acceleration. It is clear that at $t=0$,  the initial fidelity
depends on the acceleration of the teleported information, where
$\mathcal{F}^{ac}_w(0)$ is small for large values of $r_0$.
Also, it is showed that as the acceleration of all qubits in channel that based on the W-state are
increased, the fidelity decreases with minimum
bounds depends on the initial values of acceleration of the teleported information,$r_0$.

\begin{figure}[h!]
  \begin{center}
  \includegraphics[width=13pc,height=10pc]{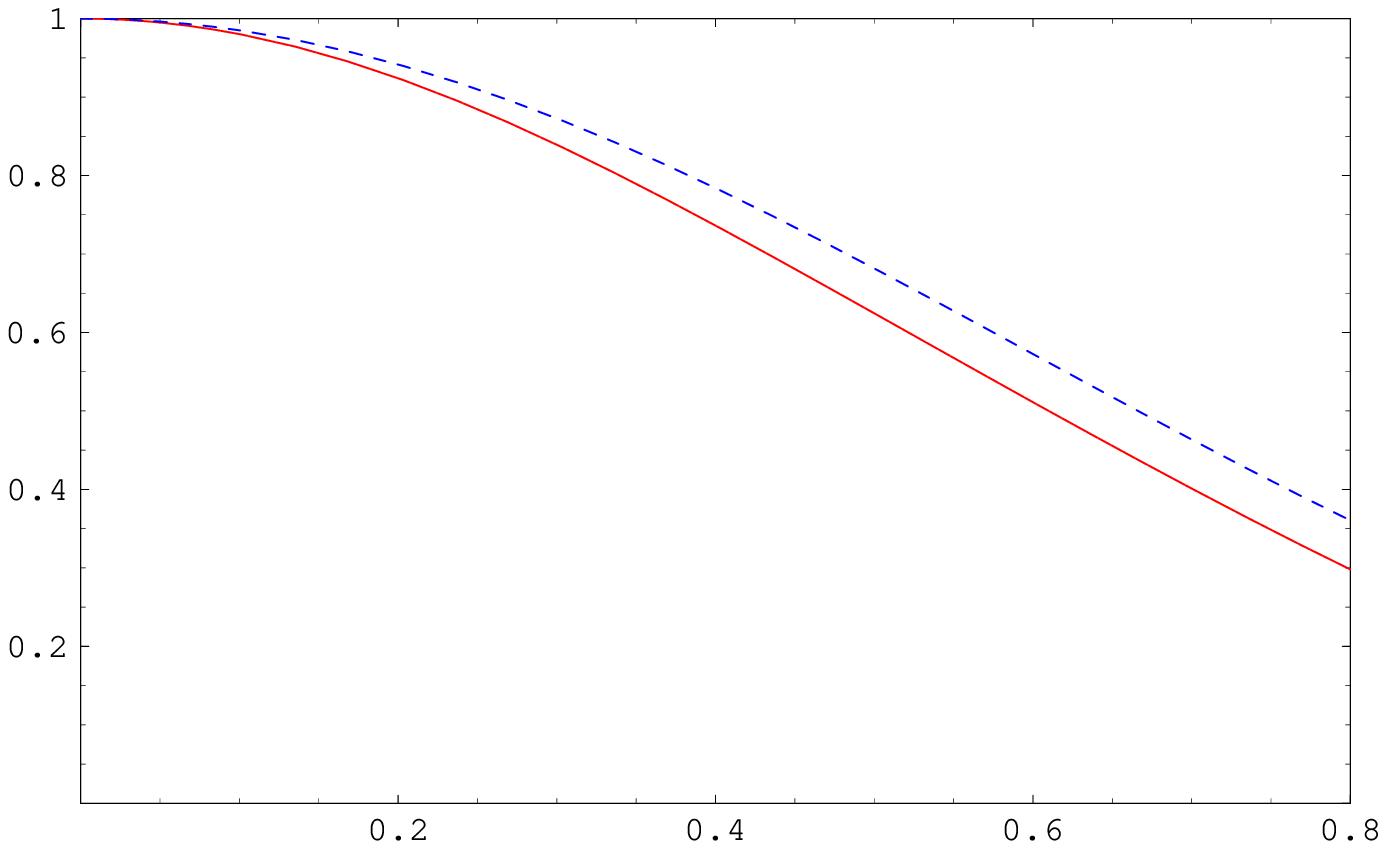}~\quad
   \includegraphics[width=15pc,height=10pc]{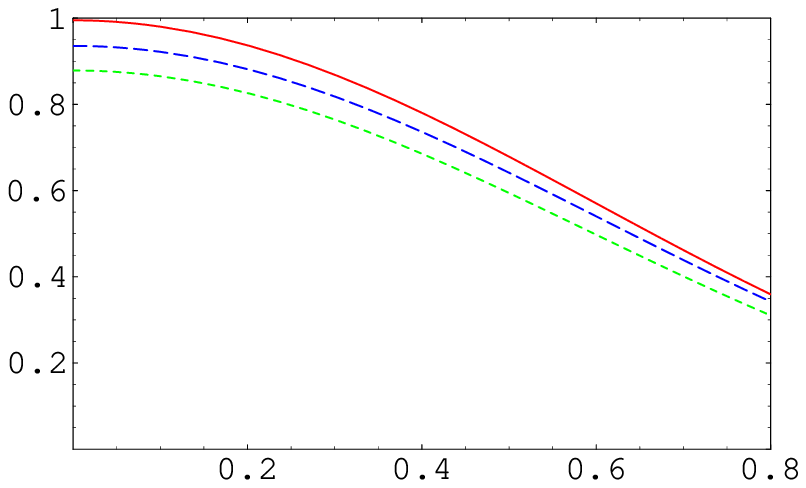}~\quad
             \put(-210,100){$(a)$} \put(-20,100){$(b)$}     \put(-360,65){$\mathcal{F}_w$}
             \put(-280,-5){$r$}
  \put(-180,65){$\mathcal{F}_{w}$}
  \put(-80,-5){$r$}
     \caption{The W-State (a)The fidelity $\mathcal{F}_{w}^{ac}$ of the accelerated (solid curve) and the fidelity $\mathcal{F}_w^{na}$ of the non-accelerated information (dot curve) and
     (b) The fidelity $\mathcal{F}_w^{ac}$ where the information acceleration $r_0=0.1,0.4,0.7$ for, the solid, dash and dot curves, respectively.}
       \end{center}
\end{figure}

\subsection{The GHZ state as a quantum channel}
In the case of using the GHZ as a quantum channel, it is assumed that the channel can be used to teleport information, either accelerated or non-accelertaed.  The three qubits of GHZ state  collaborate  together to perform the quantum
teleportation procedure as described in section 3. Similarly, after Alice performs
Bell Measurement (BM), the total state is projected into one of
the four Bell states given in Eq. (\ref{bell}). For example, if Alice
qubits are projected into $\rho_{{\psi}}^{+}$, then the two
other qubits are projected into a density operator described by
the following  $16$ elements:

\begin{eqnarray}
\varrho_{11}&=&\alpha^2C_1^2C_2^2C_3^2+\beta^2C_2^2C_3^2S_1^2,\quad
\varrho_{14}=\varrho_{41}=\alpha \beta C_1C_2C_3
\nonumber\\
\varrho_{22}&=&\alpha^2C_1^2C_2^2S_3^2+\beta^2C_2^2S_1^2S_3^2\quad
\varrho_{33}=\alpha^2C_1^2C_3^2S_2^2+\beta^2C_3^2S_1^2S_2^2, \quad
\nonumber\\
\varrho_{44}&=&\alpha^2C_1^2S_2^2S_3^2+\beta^2(1+S_1^2S_3^2S_3^2),\quad
\nonumber\\
\varrho_{12}&=&\varrho_{13}=\varrho_{21}=\varrho_{23}=\varrho_{24}=
\varrho_{31}=\varrho_{32}=
\varrho_{34}=\varrho_{42}=\varrho_{43}=0
\end{eqnarray}

Finally, Bob can end the protocol by applying adequate
operation given in Table 2 to  retrieve the teleported information with a fidelity given as,

\begin{eqnarray}\label{Gnon1}
\mathcal{F}_{g}^{na}&=& \alpha^4C_0^4C_1^2C_2^2(C_3^2 + S_3^2) +
\alpha^2\beta^2S_1^2(C_3^2 + S_3^2)+  \alpha^2\beta^2C_1^2S_2^2(C_3^2 + S_3^2)
\nonumber\\
&&+2\alpha^2\beta^2C_1C_2C_3
 \beta^4S_1^2S_2^2 (C_3^2 + S_3^2) + \beta^4.
\end{eqnarray}

\begin{table}[h]
\caption {Bob unitary operations when Charlie measures at \emph{x}-direction} \label{tab2}
\centering
\begin{tabular}{|c|c|c|}
\hline
  Alice measurement & Charlie measurement & Bob unitary operation \\
  \hline
  $\rho_{{\psi}}^{+}$ & $\emph{x}_+$ & I \\
  $\rho_{{\psi}}^{-}$ & $\emph{x}_+$ & Z \\
  $\rho_{{\phi}}^{+}$ & $\emph{x}_+$ & X \\
  $\rho_{{\phi}}^{-}$ & $\emph{x}_+$ & Y \\
  $\rho_{{\psi}}^{+}$ & $\emph{x}_-$ & Z \\
  $\rho_{{\psi}}^{-}$ & $\emph{x}_-$ & I \\
  $\rho_{{\phi}}^{+}$ & $\emph{x}_-$ & Y \\
  $\rho_{{\phi}}^{-}$ & $\emph{x}_-$ & X\\

  \hline
\end{tabular}
\end{table}

Similarly, if the GHZ state  teleports the accelerated
information, which as given in Eq. (\ref{target2}), the channel based on  GHZ state  achives the proposed procedure to retrieve the telported information at the receiver side with a fidelity takes the form:
\begin{eqnarray}\label{Gacc1}
\mathcal{F}_g^{ac}&=& \alpha^4C_0^4C_1^2C_2^2(C_3^2 + S_3^2) +
\alpha^4C_0^2S_0^2S_1^2C_2^2(C_3^2 + S_3^2)
\nonumber\\
&&+\alpha^2\beta^2C_0^2S_1^2C_2^2(C_3^2
+ S_3^2)+   2\alpha^2\beta^2C_0^2C_1C_2C_3
\nonumber\\
&&+\alpha^4S_0^2C_0^2C_1^2S_2^2(C_3^2 +
S_3^2)+\alpha^4S_0^4S_1^2S_2^2(C_3^2 + S_3^2)
\nonumber\\
&&+ \alpha^2\beta^2S_0^2S_1^2S_2^2(C_3^2 + S_3^2) + \alpha^2S_0^2(\alpha^2S_0^2 +
\beta^2)
\nonumber\\
&&+\alpha^2\beta^2C_0^2C_1^2S_2^2(C_3^2 + S_3^2)+ \alpha^2\beta^2S_0^2S_1^2S_2^2(C_3^2 + S_3^2)
\nonumber\\
&& +
 \alpha^2\beta^4S_1^2S_2^2(C_3^2 + S_3^2) + \beta^2(\alpha^2S_0^2 + \beta^2)
\end{eqnarray}
\begin{figure}[h]
  \begin{center}
  \includegraphics[width=13pc,height=10pc]{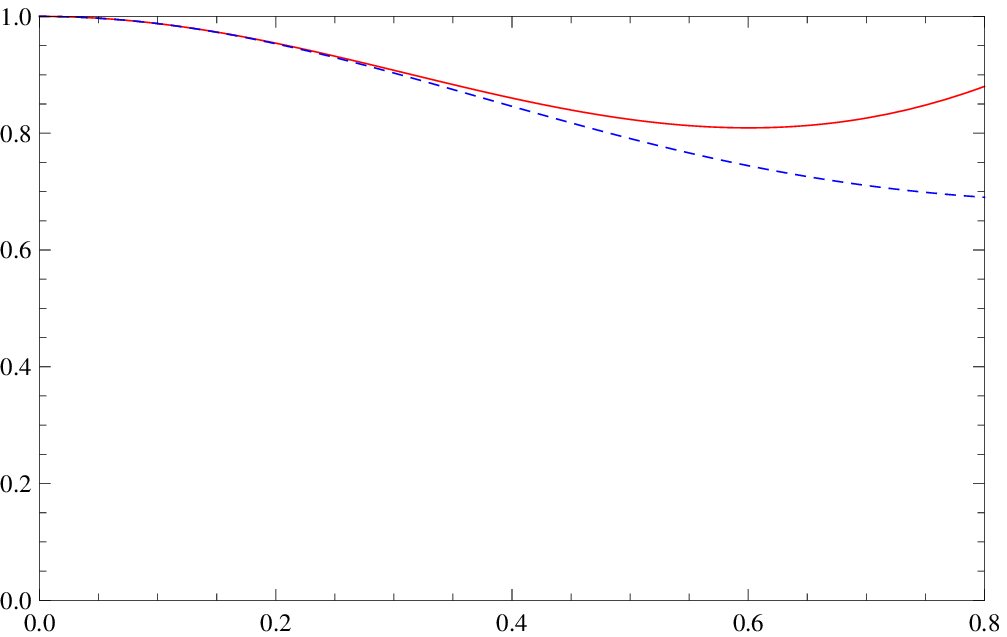}~\quad
   \includegraphics[width=15pc,height=10pc]{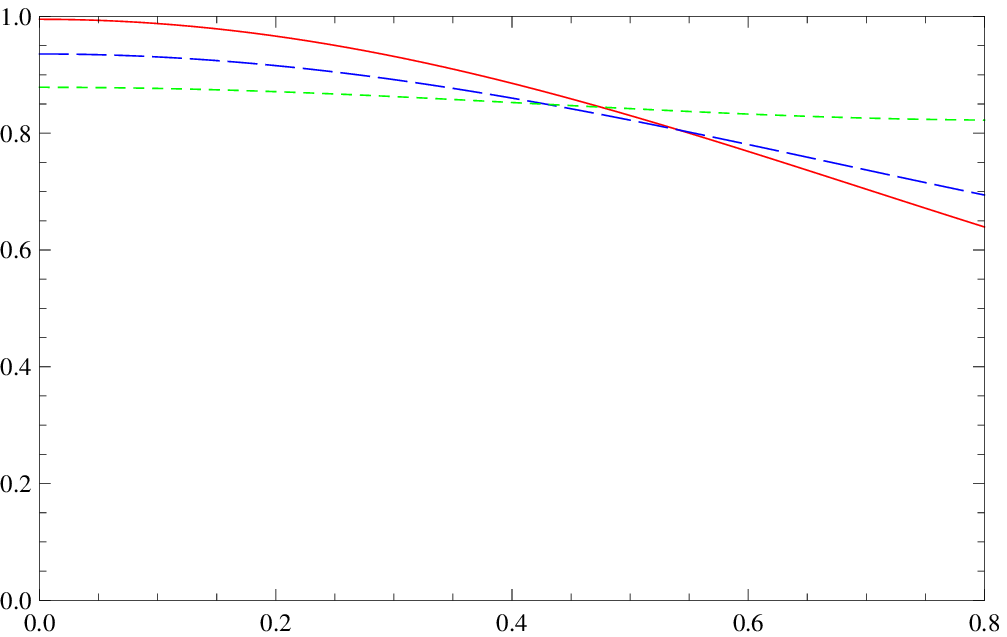}~\quad
          \put(-210,100){$(a)$} \put(-20,100){$(b)$}
      \put(-360,65){$\mathcal{F}_G$}
  \put(-280,-5){$r$}
  \put(-180,65){$\mathcal{F}_{G}$}
  \put(-80,-5){$r$}
     \caption{The GHZ-State(a)The fidelity $\mathcal{F}_{g}^{ac}$ for the accelerated (solid curve) and the fidelity $\mathcal{F}_{g}^{na}$ for the non-accelerated information (dot curve) and
     (b) The fidelity $\mathcal{F}_g^{ac}$ where the information acceleration $r_0=0.1,0.4,0.7$ for, the solid, dash and dot curves, respectively.}
       \end{center}
\end{figure}

 Figure 3 shows the behavior of the teleported information through the accelerated  GHZ state.  The panel (a),
 shows the relation between both the accelerated or non-accelerated information with
 the fidelity of each of them through the accelerated GHZ state. As
 a general remark,  the fidelities degrade as the acceleration of all qubits increase.  However, the degradation rate of the
 teleported  accelerated information is faster than
 that depicted  for the non-accelerated information. The panel (b) shows the behavior of $\mathcal{F}^{ac}_g$ for different
values of accelerated information, $r_0$. In
general, the fidelity degrades as the acceleration of the GHZ state qubits, $r_i$, increases. Also, we notice that as the difference between the values of information acceleration ($r_0$) and the values of  the qubits in GHZ state acceleration ($r_i$) decreases,
the initial fidelity tends to be larger than  that depicted for  big differences, while the minimum bound of fidelity decreases.

\subsection{The GHZ-like State as a quantum channel}
 We proceed now to the last quantum channel, which is the GHZ-like state given in Eq.(\ref{GL}. In case of teleporting non accelerated information and according to Alice measurement, the other two qubit of the channel are projected into the following density operator:

 \begin{eqnarray}
\varrho_{11}&=&\beta^2C_2^2C_3^2, \quad
\varrho_{12}=\varrho_{21}=\alpha \beta C_1C_3C_2^2,\quad
\varrho_{13}=\varrho_{31}=\alpha\beta C_1C_2C_3^2
 \nonumber\\
\varrho_{14}&=&\varrho_{41}=\beta^2 C_2C_3,\quad
\varrho_{22}=\alpha^2 C_1^2C_2^2+\beta^2 C_2^2S_1^2+\beta^2
C_2^2S_3^2
 \nonumber\\
 \varrho_{23}&=&\alpha^2 C_2C_3C_1^2+\beta^2
C_3^2S_1^2+\beta^2 C_2C_3S_1^2, \quad
\varrho_{24}=\varrho_{42}=(\alpha \beta C_0C_1C_2(1+S_3^2)
\nonumber\\
 \varrho_{32}&=&\alpha^2  C_0^2C_1^2C_2C_3+(\alpha^2
S_0^2 +\beta^2)C_2C_3S_1^2,
\nonumber\\
 \varrho_{33}&=&\alpha^2
C_0^2C_1^2C_3^2+(\alpha^2 S_0^2 +\beta^2)C_3^2S_1^2+C_3^2S_2^2
\nonumber\\
\varrho_{34}&=&\varrho_{43}=(\alpha \beta
C_1C_3(1+S_2^2),\quad
 \varrho_{43}=(\alpha \beta
C_0C_1C_3(1+S_2^2)
\nonumber\\
\varrho_{44}&=&(\alpha^2
C_1^2(S_2^2+S_3^2)+\beta^2S_1^2(S_2^2+S_3^2)+\beta^2(1+S2_2S3^2).
\end{eqnarray}

 Bob ends the procedure after he performs an appropriate operation from those given in Table 3 with a fidelity depending on Alice and Charlie measurements. Here we have two different cases for example :

\begin{itemize}
\item If Alice measures $\rho_{{\phi}}^{+}$ and Charlie
measures "0", then the fidelity of the teleported information is given as
\begin{eqnarray}\label{GLnon30}
F_{gl}^{ac}&=& \alpha^4C_2^2C_3^2 + \alpha^2\beta^2C_3^2(S1^2+S_2^2)
\nonumber\\
&&+\beta^4C_1^2C_3^2
 + 2\alpha^2\beta^2C_1C_2C_3^2
\end{eqnarray}

\item If Alice measures $\rho_{{\phi}}^{+}$ and Charlie
measures "1", then the fidelity of the teleported information is given as
\begin{eqnarray}\label{GLnon31}
F_{gl}^{ac}&=& \alpha^4S_1^2 (S_2^2 + S_3^2)+ \alpha^2\beta^2C_1^2(S_2^2 + S_3^2)+\alpha^2\beta^2C_2^2(S_1^2 + S_3^2)
\nonumber\\
&& +\alpha^4(S_2^2S_3^2 + 1)+\beta^4C_1^2C_2^2 + 2\alpha^2\beta^2C_1C_2(1+S_3^2)
 \end{eqnarray}
\end{itemize}

\begin{table}[h]
\caption {Bob unitary operations when Charlie measures $"0"$ and $"1"$ } \label{tab3}
\centering
\begin{tabular}{|c|c|c|}
\hline
  Alice measurement & Charlie measurement & Bob unitary operation \\
  \hline
  $\rho_{{\psi}}^{+}$ & {0}& X \\
  $\rho_{{\psi}}^{-}$ & {0} & Y \\
  $\rho_{{\phi}}^{+}$ & {0} & I \\
  $\rho_{{\phi}}^{-}$ & {0} & Z \\
  $\rho_{{\psi}}^{+}$ & {1}& I \\
  $\rho_{{\psi}}^{+}$ & {1}& Z \\
  $\rho_{{\phi}}^{+}$ & {1}& X \\
  $\rho_{{\phi}}^{+}$ & {1}& Y \\
  \hline
\end{tabular}
\end{table}

On the other hand, if the teleported information is accelerated through the GHZ-like state, the teleported information at Bob side will have a fidelity depends on  Alice and Charlie measurements ,for example:

\begin{itemize}

\item If Alice measures $\rho_{{\phi}}^{+}$ and Charlie
masures "0", then the fidelity of the teleported state is given as:
\begin{eqnarray}\label{GLacc30}
\mathcal{F}_{gl}^{na}&=& (\beta^2C_1^2C_3^2(\alpha^2S_0^2 + \beta^2) +
\alpha^2S_0^2C_1^2C_3^2(\alpha^2S_0^2 + \beta^2)
\nonumber\\
&&+ \alpha^4C_0^2S_0^2C_3^2(S_1^2 +
S_2^2)+\alpha^2\beta^2C_3^2(C_0^2S_1^2 + S_2^2)
\nonumber\\
&& + \alpha^4C_0^4C_2^2C_3^2 +
2\alpha^2\beta^2C_0^2C_1C_2C_3^2)
\end{eqnarray}

\item If Alice measures $\rho_{{\phi}}^{+}$ and Charlie
measures "1", then the fidelity of the teleported state is given as:
\begin{eqnarray}\label{GLacc31}
\mathcal{F}_{gl}^{na}&=& \alpha^2C_0^2C_1^2S_3^2(\alpha^2S_0^2 + \beta^2) +
\alpha^2C_0^2C_1^2S_2^2(\alpha^2S_0^2 + \beta^2)
\nonumber\\
&&+ \alpha^2S_0^2C_1^2C_2^2(\alpha^2S_0^2 +
\beta^2)+  \beta^2C_1^2C_2^2(\alpha^2S_0^2 + \beta^2)
\nonumber\\
&&+ \alpha^4S_0^2C_0^2C_2^2(S_1^2 +
S3^2) + \alpha^2\beta^2C_0^2C_2^2(S_1^2 + S_3^2)
\nonumber\\
&&+ 2\alpha^2\beta^2C_0^2C_1C_2 + 2\alpha^2\beta^2C_0^2C_1C_2S_3^2
\nonumber\\
&&+\alpha^4C_0^4(S_1^2S_2^2 + S_1^2S_3^2 + S_2^2S_3^2 + 1)
\end{eqnarray}
\end{itemize}

\begin{figure}[h]
  \begin{center}
  \includegraphics[width=13pc,height=10pc]{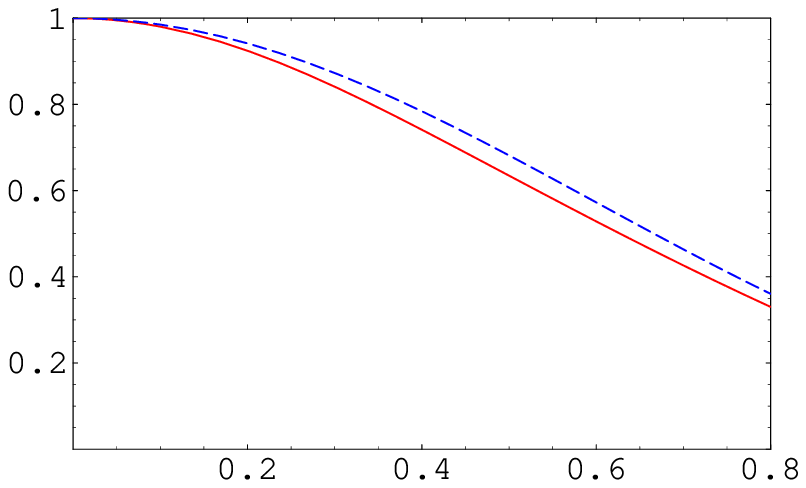}~\quad
  \includegraphics[width=15pc,height=10pc]{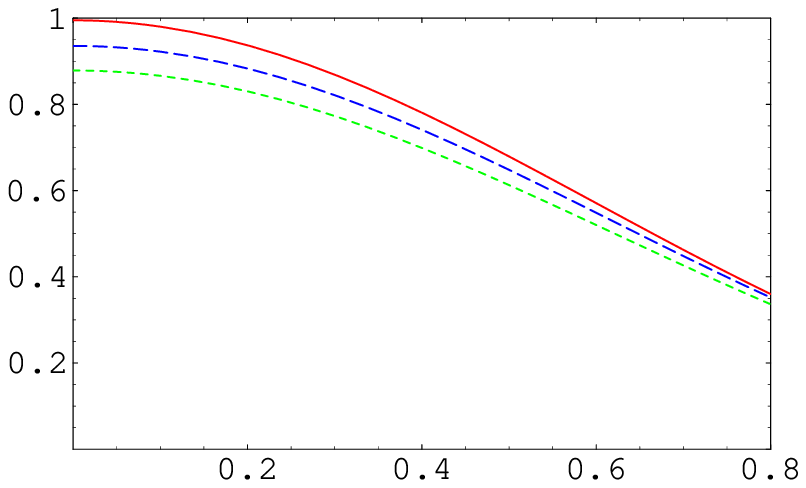}~\quad
            \put(-210,100){$(a)$} \put(-20,100){$(b)$}
      \put(-360,65){$\mathcal{F}_{GL}$}
  \put(-280,-5){$r$}
  \put(-180,65){$\mathcal{F}_{GL}$}
  \put(-80,-5){$r$}
     \caption{The GHZ-like state(a)The fidelity $\mathcal{F}_{gl}^{ac}$ of the accelerated information when Alice projects the
     system into $\rho_{psi}^+$ (solid curve) and the The fidelity $\mathcal{F}_{gl}^{na}$ of the non-accelerated information (dot curve), both when Charlie measures $"0"$.
     (b) The fidelity $\mathcal{F}_{gl}^{ac}$ where $r_0=0.1,0.4,0.7$ for, the solid, dash and dot curves, respectively.}
       \end{center}
\end{figure}

The fidelity of the teleported information through the GHZ-like state is depicted in Figure 4.
It is  shown that the fidelity degrades as the acceleration of the  qubits of the channel based on GHZ-like state increases. Also, the degradation rate depends on the measurements performed by the qubit of GHZ-like state.
Panel (a) shows the
behavior of the fidelity of information, either accelerated or non-accelerated, in the situation that Alice performs Bell measurement using $\phi^{+}$ and Charlie measures "$0"$. It is clear that the degradation rate
 of the fidelity  for accelerated information is  faster than that depicted for the
 non-accelerated information, same performance in case of the W-state.
The panel (b) shows the behavior of $\mathcal{F}^{ac}_{gl}$ for different
values of accelerated information, $r_0$, where the fidelity degrades as the acceleration of all the qubit of the  channel based on GHZ-like, $r_i$, increases.

\section{Conclusion and Future Work}

In this paper, we investigated the possibility of using
multi-accelerated qubits as a quantum channel to perform quantum
teleportation. Namely, These  states are the W, the GHZ and the
GHZ-like states. In general, we found that the fidelity of the
teleported information degrades as the acceleration of the qubits
of the used states increases.

It is shown in the paper that, the degradation rate of the
teleported state is large for accelerated information in the case
of using the W state and the GHZ-like state as a quantum channel where it is
small in the case of using the GHZ state as a quantum channel. This
certifies that the rate of fidelity degradation depends on the
used channel. The comparison among the three channels showed that
the GHZ state is  the optimum, comparing to the W-state and
GHZ-like state, for teleporting information either accelerated or
non-accelerated.

In addition, the paper investigated the effect of different values
of the accelerated information through each accelerated state.
We found that, the initial fidelity of information decreases at
high values of acceleration of the information. The minimum value,
which the fidelity approaches, depends on the used state and
acceleration rate of the information. However, the minimum values
of the fidelities through the W and GHZ-like states is smaller
than that depicted in the case of  the GHZ state.

Overall, the contribution of the paper is concluded in the following two items:
\begin{itemize}
   \item The three  accelerated states: the W, the GHZ and
the GHZ-like state can be used as accelerated quantum channel to
teleport accelerated or non-accelerated information.
   \item The fidelity of teleporting accelerated or non-accelerated information through the
 accelerated GHZ state as quantum channel is much better than either the W or the GHZ-like  states.
 \end{itemize}

Using the proposed teleportation procedure in quantum communication applications is one of our future target.
 We are looking forward to develop a prototype for communication via accelerated entangled quantum network
  written by a high-level programming language.


\begin{thebibliography}{nas}
\bibitem{Horodecki}
R. Horodecki, P. Horodecki, M. Horodecki, and K. Horodecki, {\it Quantum
Entanglement,} Reviews of Modern Physics, 81, 2,  865-
942, 2009.

\bibitem{Bennett1}
C. Bennett and  G. Brassard , {\it Quantum Cryptography: Public Key Distribution and Coin Tossing}, in Proceedings of IEEE International Conference on Computers, Systems, and Signal Processing, Bangalore, India, 1984.


\bibitem{RIEFFEL}
E. Rieffel and W. Polak ,{\it An Introduction to Quantum Computing for Non-Physicists}, ACM Computing Surveys, 32, 3, 300-335, 2000.

\bibitem{Metwally2004}
N. Metwally, M. Abdelaty and A.-S.F. Obada {\it Quantum
Teleportation via Entangled States Generated by the Jaynes-
Cummings Model,} Chaos, Solitons and Fractals, 22, 529, 2004.

\bibitem{Metwally2009}
N. Metwally,{\it Entanglement and information  transfer protocol between two qubits,} International Journal of Modern Physics C, 20, 769-780, 2009.

\bibitem{Ursin}
R. Ursin , T. Jennewein , and M. Aspelmeyer, {\it Communications: Quantum
Teleportation across the Danube,} Nature, 430: 849, 2004.





\bibitem{Gorbachev}
V.N. Gorbachev and A.I Trubilko,
{\it Teleportation and Dense Coding via a Multi-Particle Quantum Channel of the GHZ-Class
,} Sov. Phys. JETP, 91, 894, 2000.

\bibitem{Joo}
J. Joo, Y-J. Park, S.Oh and J. Kim, {\it Quantum Teleportation via a W State,} New J. Phys. 5, 136, 2003.

\bibitem{Kan}
K. Yang, L. Huang, W. Yang, F. Song ,  {\it Quantum Teleportation via GHZ-like State,} International Journal of Theoretical Physics, 48, 2, 516-521, 2009.

\bibitem{Bennett2}
C. Bennett, G. Brassard , C. Crepeau , R. Jozsa , A. Peres  and
W. Wootters , {\it Teleporting an Unknown Quantum State via Dual Classical and Einstein-Podolsky-Rosen Channels,} Phys. Rev. Lett., 70, 1895, 1993.


\bibitem{Zeilinger}
D. Bouwmeester, J-W. Pan, M. Mattle, H. Weinfurter  and
A. Zeilinger, {\it Experimental Quantum Teleportation,} Nature, 390, 575, 1997.

\bibitem{Zeilinger2}
X-S. Ma, T. Herbst, T. Scheidl, D. Wang, S. Kropatschek, W. Naylor, B. Wittmann, A. Mech, J. Kofler, E. Anisimova, V. Makarov, T. Jennewein, R. Ursin and A. Zeilinger, {\it Quantum Teleportation over 143 Kilometres using Active Feed-Forward,} Nature, 489, 269, 2012.

\bibitem{Ivan}
I. Capraro, A. Tomaello, A. Dall’Arche, F. Gerlin, R. Ursin,
G. Vallone, and P. Villores, {\it Impact of Turbulence in Long Range Quantum and Classical Communications,} Physical Review Letters, 109, 200502, 2012

\bibitem{Jeongwan}
J. Jin, S. Park and Y. Kwon, {\it Quantum Teleportation in Three Parties with an Accelerated Receiver,},
Chaos, Solitons and Fractals, 28,  2, pp 313-319, 2006.

\bibitem{Karlsson}
A. Karlsson and M. Bourennane,{\it Quantum Teleportation Using Three-Particle Entanglement,} Phys. Rev. A, 58, pp 4394–4400, 1998.

\bibitem{Alsing1}
P. M. Alsing,{\it Teleportation in a Non-Inertial Frame,} Journal of Optics B: Quantum and Semiclassical Optics, 6, S834, 2004.

\bibitem{Alsing2}
P. M. Alsing and G. J. Milburn, {\it Teleportation with a Uniformly Accelerated Partner,} Phys. Rev. Lett, 91, 18, 2003.

\bibitem{Metwally1}
N. Metwally, {\it Teleportation of Accelerated Information,} Journal of the Optical Society of America B, 30, 233-237, 2013.

\bibitem{Metwally2}
N. Metwally, {\it Usefulness Classes of Travelling Entangled
Channels in Non-Intertial Frames,} International Journal of Modern Physics B, 27, 18, 2013.






\bibitem{Martın}
E. Martínez and J. León, {\it Fermionic entanglement that survives a black hole,}  Phys. Rev. A, 80, 012314, 2009





\bibitem{Alaa} A. Sagheer and H. Hamdoun, {\it Dynamics of Multi-Qubit States in Non-Inertial Frames for Quantum Communication Applications,} Quantum Information and Computation Journal, 14, 255-264, 2014.

    \end{thebibliography}
\end{document}